\documentclass{emulateapj}
\usepackage{color}

\newcommand{\kms}{\,km~s$^{-1}$}      

\def\h50{\, h_{50}^{-1}}

\def\spose#1{\hbox to 0pt{#1\hss}}
\def\simlt{\mathrel{\spose{\lower 3pt\hbox{$\mathchar"218$}}
     \raise 2.0pt\hbox{$\mathchar"13C$}}}
\def\simgt{\mathrel{\spose{\lower 3pt\hbox{$\mathchar"218$}}
     \raise 2.0pt\hbox{$\mathchar"13E$}}}

\newcommand{\myemail}{eros.vanzella@oabo.inaf.it}

\newcommand{\hii}{\textrm{H}\textsc{ii}}

\newcommand{\oiidoub}{[\textrm{O}\textsc{ii}]\ensuremath{\lambda3727,3729}}

\newcommand{\oiiiv}{[\textrm{O}\textsc{iii}]\ensuremath{\lambda5007}}

\newcommand{\oiiidoub}{[\textrm{O}~\textsc{iii}]\ensuremath{\lambda\lambda4959,5007}}
 
\newcommand{\ha}{\ifmmode {\rm H}\alpha \else H$\alpha$\fi}
\newcommand{\hb}{\ifmmode {\rm H}\beta \else H$\beta$\fi}
\newcommand{\lya}{\ifmmode {\rm Ly}\alpha \else Ly$\alpha$\fi}
\newcommand{\pg}{\ifmmode {\rm P}\gamma \else Pa$\gamma$\fi}
\newcommand{\lyb}{\ifmmode {\rm Ly}\beta \else Ly$\beta$\fi}
\newcommand{\lyg}{\ifmmode {\rm Ly}\gamma \else Ly$\gamma$\fi}

\newcommand{\ciiidoub}{\textrm{C}\textsc{iii}]\ensuremath{\lambda\lambda1907,1909}}

\newcommand{\civ}{\textrm{C}\textsc{iv}\ensuremath{\lambda1548,1550}}
\newcommand{\civalone}{\textrm{C}\textsc{iv}}

\newcommand{\civblue}{\textrm{C}\textsc{iv}\ensuremath{\lambda 1548}}
\newcommand{\oiiiuvred}{\textrm{O}\textsc{iii}]\ensuremath{\lambda1666}}
\newcommand{\heii}{\textrm{He}\textsc{ii}\ensuremath{\lambda1640}}
\newcommand{\oiiiuv}{\textrm{O}\textsc{iii}]\ensuremath{\lambda1661,1666}}

\def\kms{km s$^{-1}$}

\def\erg{erg s$^{-1}$ cm$^{-2}$ \AA$^{-1}$}
\def\ergs{\ifmmode \mathrm{erg\hspace{1mm}s}^{-1} \else erg s$^{-1}$\fi}
\def\ergscm{erg s$^{-1}$ cm$^{-2}$}
\def\lumi{erg s$^{-1}$}
\def\micron{\ifmmode \mu\mathrm{m} \else $\mu$m\fi}
\def\msun{\ifmmode \mathrm{M}_{\odot} \else M$_{\odot}$\fi}
\def\msunyr{\ifmmode \mathrm{M}_{\odot} \hspace{1mm}{\rm yr}^{-1} \else $\mathrm{M}_{\odot}$ yr$^{-1}$\fi}
\def\zsun{\ifmmode Z_{\odot} \else Z$_{\odot}$\fi}
\def\lsun{\ifmmode L_{\odot} \else L$_{\odot}$\fi}
\def\mstar{\ifmmode \mathrm{M}_{\star} \else M$_{\star}$\fi}

\shorttitle{Resolving parsec-scale star-forming regions at z=3.22}
\shortauthors{Vanzella et al.}

\begin{document}

\title{Magnifying the early episodes of star formation: super star clusters at cosmological distances}

\author{\sc E. Vanzella\altaffilmark{1,*},
M. Castellano\altaffilmark{2},
M. Meneghetti\altaffilmark{1},
A. Mercurio\altaffilmark{3},
G. B. Caminha\altaffilmark{4},
G. Cupani\altaffilmark{5},
F. Calura\altaffilmark{1},
L. Christensen\altaffilmark{6},
E. Merlin\altaffilmark{2},
P. Rosati\altaffilmark{4},
M. Gronke\altaffilmark{7},
M. Dijkstra\altaffilmark{7},
M. Mignoli\altaffilmark{1},
R. Gilli\altaffilmark{1},
S. De Barros\altaffilmark{8},
K. Caputi\altaffilmark{9},
C. Grillo\altaffilmark{6,10},
I. Balestra\altaffilmark{11},
S. Cristiani\altaffilmark{5},
M. Nonino\altaffilmark{5},
E. Giallongo\altaffilmark{2},
A. Grazian\altaffilmark{2},
L. Pentericci\altaffilmark{2},
A. Fontana\altaffilmark{2},
A. Comastri\altaffilmark{1},
C. Vignali\altaffilmark{12,1},
G. Zamorani\altaffilmark{1},
M. Brusa\altaffilmark{12,1},
P. Bergamini\altaffilmark{13},
P. Tozzi\altaffilmark{14}
}

\altaffiltext{1}{INAF--Osservatorio Astronomico di Bologna, via Gobetti 93/3, 40129 Bologna, Italy}
\altaffiltext{2}{INAF--Osservatorio Astronomico di Roma, via Frascati 33, 00040 Monteporzio, Italy}
\altaffiltext{3}{INAF -- Osservatorio Astronomico di Capodimonte, Via Moiariello 16, I-80131 Napoli, Italy}
\altaffiltext{4}{Dipartimento di Fisica e Scienze della Terra, Universit\`a di Ferrara, via Saragat 1, 44122 Ferrara, Italy}
\altaffiltext{5}{INAF - Osservatorio Astronomico di Trieste, via G. B. Tiepolo 11, I-34131, Trieste, Italy}
\altaffiltext{6}{Dark Cosmology Centre, Niels Bohr Institute, University of Copenhagen, Juliane Maries Vej 30, DK-2100 Copenhagen, Denmark}
\altaffiltext{7}{Institute of Theoretical Astrophysics, University of Oslo, Postboks 1029 Blindern, NO-0315 Oslo, Norway}
\altaffiltext{8}{Observatoire de Gen\`eve, Universit\'e de Gen\`eve, 51 Ch. des Maillettes, 1290, Versoix, Switzerland}
\altaffiltext{9}{Kapteyn Astronomical Institute, University of Groningen, Postbus 800, 9700 AV, Groningen, The Netherlands}
\altaffiltext{10}{Dipartimento di Fisica, Universit\`a  degli Studi di Milano, via Celoria 16, I-20133 Milano, Italy}
\altaffiltext{11}{University Observatory Munich, Scheinerstrasse 1, D-819 M\"unchen, Germany}
\altaffiltext{12}{Dipartimento di Fisica e Astronomia, Universit\`a degli Studi di Bologna, Viale Berti Pichat 6/2, 40127 Bologna, Italy}
\altaffiltext{13}{Dipartimento di Fisica e Astronomia ``G. Galilei'', Universit\`a di Padova, Vicolo dell'Osservatorio 3, I-35122, Italy}
\altaffiltext{14}{INAF -- Osservatorio Astrofisico di Arcetri, Largo E. Fermi, I-50125, Firenze, Italy}

\altaffiltext{*}{\myemail}
\altaffiltext{$\dagger$}{Based on observations collected at the European Southern Observatory for Astronomical research in the
Southern Hemisphere under ESO programmes P095.A-0840, P095.A-0653, P186.A-0798.}

\begin{abstract}
  We study the spectrophotometric properties of a highly magnified ($\mu \simeq 40-70$)
  pair of stellar systems identified at z=3.2222 behind the Hubble Frontier Field galaxy
  cluster MACS~J0416. Five multiple images (out of six) have been spectroscopically
  confirmed by means of VLT/MUSE and VLT/X-Shooter observations. Each image includes
  two faint ($m_{UV} \simeq 30.6$), young ($\lesssim 100$~Myr), low-mass ($<10^{7}$\msun), 
  low-metallicity (12+Log(O/H)$\simeq$7.7, or 1/10 solar) and compact (30 pc effective radius)
  stellar systems separated by $\simeq 300$ pc, after correcting for lensing
  amplification. We measured several rest-frame ultraviolet and optical narrow
  ($\sigma_v \lesssim 25$ \kms) high-ionization lines.
  These features may be the signature of very hot ($T>50000$ K) stars within dense stellar clusters, 
  whose dynamical mass is likely dominated by the stellar component. 
  Remarkably, the ultraviolet metal lines are not accompanied by \lya\ emission (e.g., \civalone\ / \lya\ $> 15$),
  despite the fact that the \lya\ line flux is expected to be 150 times brighter (inferred from the \hb\ flux).
  A spatially-offset, strongly-magnified ($\mu>50$) \lya\ emission with a spatial extent $\lesssim7.6$~kpc$^2$
  is instead identified 2~kpc away from the system. The origin of such a faint emission can be 
  the result of fluorescent \lya\ induced by a transverse leakage of ionizing radiation emerging
  from the stellar systems and/or can be associated to an underlying and barely detected object
  (with $m_{UV}>34$ de-lensed). This is the first confirmed metal-line emitter at such low-luminosity
  and redshift without \lya\ emission, suggesting that, at least in some cases, a non-uniform covering
  factor of the neutral gas might hamper the \lya\ detection. 
    \end{abstract}

\keywords{cosmology: observations --- galaxies: formation}

\section{Introduction}
The investigation of the ionizing properties of  young, low mass star-forming systems
caught at $z\simeq 3$, i.e., nearly 1 Gyr after reionization ended and the possible analogy with similar systems identified during
reionization ($z>6$) represents today a strategic line of research, especially at poorly explored, low-luminosity regimes \citep[e.g.,][]{amorin17}. 
The detection of nebular high ionization emission lines 
has underscored the considerable contribution of hot and massive stars to the transparency of the medium in low-luminosity ($L << L^*$) systems 
\citep[][]{vanz16a}, as expected in scenarios in which stellar winds and supernova explosions blow cavities in the interstellar
medium  \citep[e.g.,][]{jaskot13,micheva17}. 
In fact, the rest-frame ultraviolet and optical line ratios observed in $z\simeq 0$,  $z=2-3$ and $z>6$ galaxies
suggest a possible evolution of the ionizing radiation field, such that at high redshift the presence of hot stars seems more common,
with blackbody mean effective temperatures of the order of 50.000--60.000 K at $z=2-3$ \citep{fosbury03,steidel14,holden16},
or even hotter at $z>6$ \citep{stark15, mainali17}. \\
Moreover, in very recent times, strong gravitational lensing has often been used to investigate the
physical properties of such intrinsically faint, low-mass systems. 
Its use is now reaching its maturity thanks to dedicated deep imaging studies 
in lensed fields (e.g., the Hubble Frontier Fields program, HFF hereafter,  \citealt{lotz14,lotz16, koekemoer14}), 
as performed on $3<z<8$ dropout galaxies \citep[e.g.,][]{bouwens17} 
and spectroscopic surveys \citep[e.g.,][]{vanz14,stark14,karman17, cam16c, stark17,mainali17, vanz16a, vanz17b}. 
In this regard, the exceptional line flux sensitivity of the integral field spectrograph MUSE
mounted on the VLT 
\citep[][]{bacon10}\footnote{{\it www.eso.org/sci/facilities/develop/instruments/muse.html}} has driven considerable progress. 
In fact, MUSE allows the identification and characterization 
of extremely faint (and small) line emitters at $3<z<6.6$, without the need of specific pre-selection of
the targets in a relatively large field of view ($1'\times1'$), aptly matching strongly magnified regions
of the sky. In these investigations, 
strong lensing turned out to be essential for two reasons: (1) before the completion of the 
Extremely Large Telescope (ELT), it is not possible to spatially resolve sources
with effective radii lower than 150 pc ($<20$mas) at $z>3$ in non-lensed fields.
Conversely, galaxy cluster lensing has already enabled us to perform light profile
fitting of dwarf and super star-clusters with radii $15-100$ pc at
$3<z<6.5$ \citep[e.g.,][]{vanz16a, vanz17b, bouwens17}; 
(2) the increased S/N of the SEDs and emission lines
allows us to access the physical properties down to (intrinsic) magnitudes $m>29$ and
line fluxes of $>5-10\times 10^{-20}$ cgs.

MUSE observations on HFFs are leading us to the construction of a reference sample at $z<4$ of faint, star-forming compact analogues 
of the systems active during cosmic reionization. Such a task is particularly important, 
given the imminent launch of the James Webb Space Telescope (JWST), whose primary aim is to capture the first episodes of star formation at $z>6$.\\ 
At $z\simeq 2-3$ the optical rest-frame is accessible from ground-based spectrographs and a large number
of strongly lensed systems have been identified \citep[e.g.,][]{alavi16, cam16a,cam16c,cam16b,karman17}, 
allowing us to probe spectroscopically extremely low-luminosity and small mass regimes with unprecedented
detail \citep[e.g.,][]{christ12, stark14, vanz16a}. 
Some of these systems have already provided a valuable insight of the  physical conditions characterising newborn,  
highly-ionized, dwarf galaxies  \citep[e.g.,][]{vanz16a, vanz16b} and extremely compact star clusters, presenting features very similar 
to those expected during the formation of globular clusters \citep{vanz17b}. 
A new intriguing system of this kind is that discussed here, i.e.,
a strongly lensed system at $z=3.2222$ which shows high ionization lines detected behind the
HFF galaxy cluster MACS~J0416 \citep[][]{cam16c}. 
Beside its compactness and its  extreme faintness, the distinguishing feature of this system is the presence of strong ionization lines which 
contrasts with the absence of \lya\ emission. These features are very rare among known local and distant star-forming systems
\citep[e.g.,][]{stark14, stark15, henry15, mainali17, shibuya17, smit17}. 
The intriguing nature of our system and of its ionization structure 
provides us with an insightful example of the geometrical complexity of high-redshift young, ionized systems. 
The energy and the momentum continuously deposited by massive, young stellar associations can lead to puzzling structures 
such as our object, which also represents a valuable benchmark for interstellar photo-ionization models. \\
In this work, we use the Hubble Frontier Fields HST-dataset\footnote{www.stsci.edu/hst/campaigns/frontier-fields/HST-Survey}, 
in particular three optical and four near-infrared bands: F435W, F606W, F814W (HST/ACS) and 
F105W, F125W, F140W and F160W (HST/WFC3),
with typical limiting AB magnitudes 28.5 - 29.0 calculated at $5\sigma$ depth. In addition to the
seven HST bands, the publicly available Hawk-I@VLT $K$s band and the Spitzer/IRAC 3.6 and 
4.6 $\mu$m data have also been  included (the description of the
data-set, the extraction of the PSF-matched photometry and the construction of the photometric
catalog are provided in \citealt{castellano16b} and \citealt{merlin16}). 

This paper is organized as follows. 
In Sect. 2, a brief description of the multiple-image magnification pattern of our system is presented. 
Our main results are presented in Sect.~3, and discussed in Sect. 4. 
Finally, in Sect. 5 some conclusions are drawn.

Throughout the paper, we assume a flat cosmology with $\Omega_{M}$= 0.3,
$\Omega_{\Lambda}$= 0.7 and $H_{0} = 70$ km s$^{-1}$ Mpc$^{-1}$.

\begin{figure*}
 \epsscale{1.0}
 \plotone{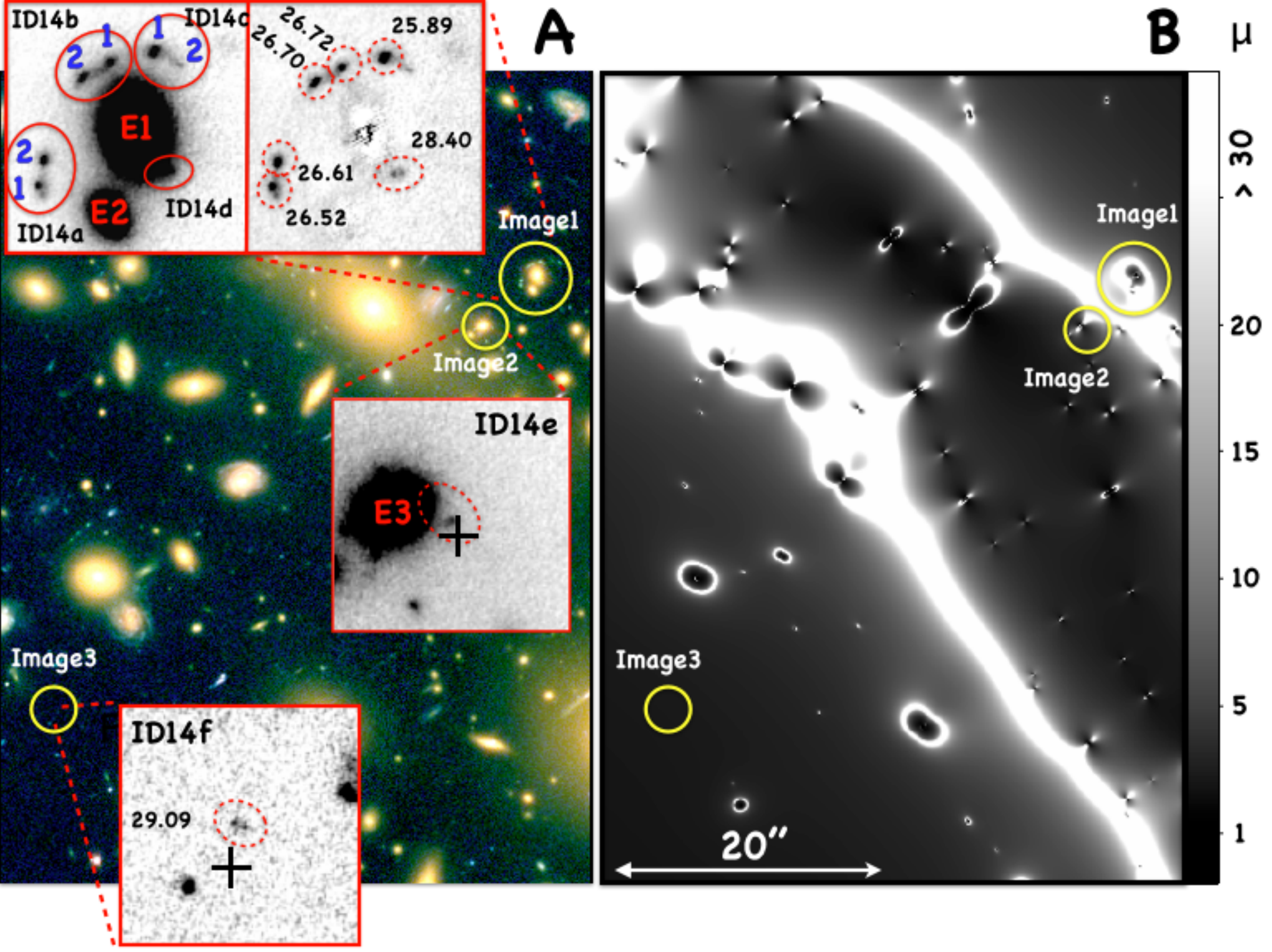} 
\caption{Panel A: the HST color image of the galaxy cluster MACS~J0416 showing the position of the three multiple images
expected at the redshift of ID14 generated by the galaxy cluster (``Image1'', ``Image2'' and ``Image3''). ``Image1''
is further magnified by the elliptical galaxy cluster members E1+E2 into four additional multiple images (ID14a,b,c,d). 
The left and right top insets show the F814W image with and without the subtraction of the galaxies E1+E2.
Each image of ID14 contains two further components, ``1'' and ``2''. Magnitudes correspond to the average
values obtained from all the HST bands (except F435W).
The other  two multiple images ``Image2'' (ID14e) and ``Image3'' (ID14f) are also shown in the insets (red dotted circles),
where the black crosses mark the position predicted by the lens model. All insets have a size of $3.3''$.
Panel B: magnification 
map ($\mu$, coded according to the
greyscale bar on the right) derived from the lens model by \citet[][]{cam16c}:  
large magnifications are expected close to the critical lines, between ``Image1'' and ``Image2'' 
and between ID14a,b and c.
\label{pano}}
\end{figure*}

\begin{figure}
 \epsscale{1.0}
 \plotone{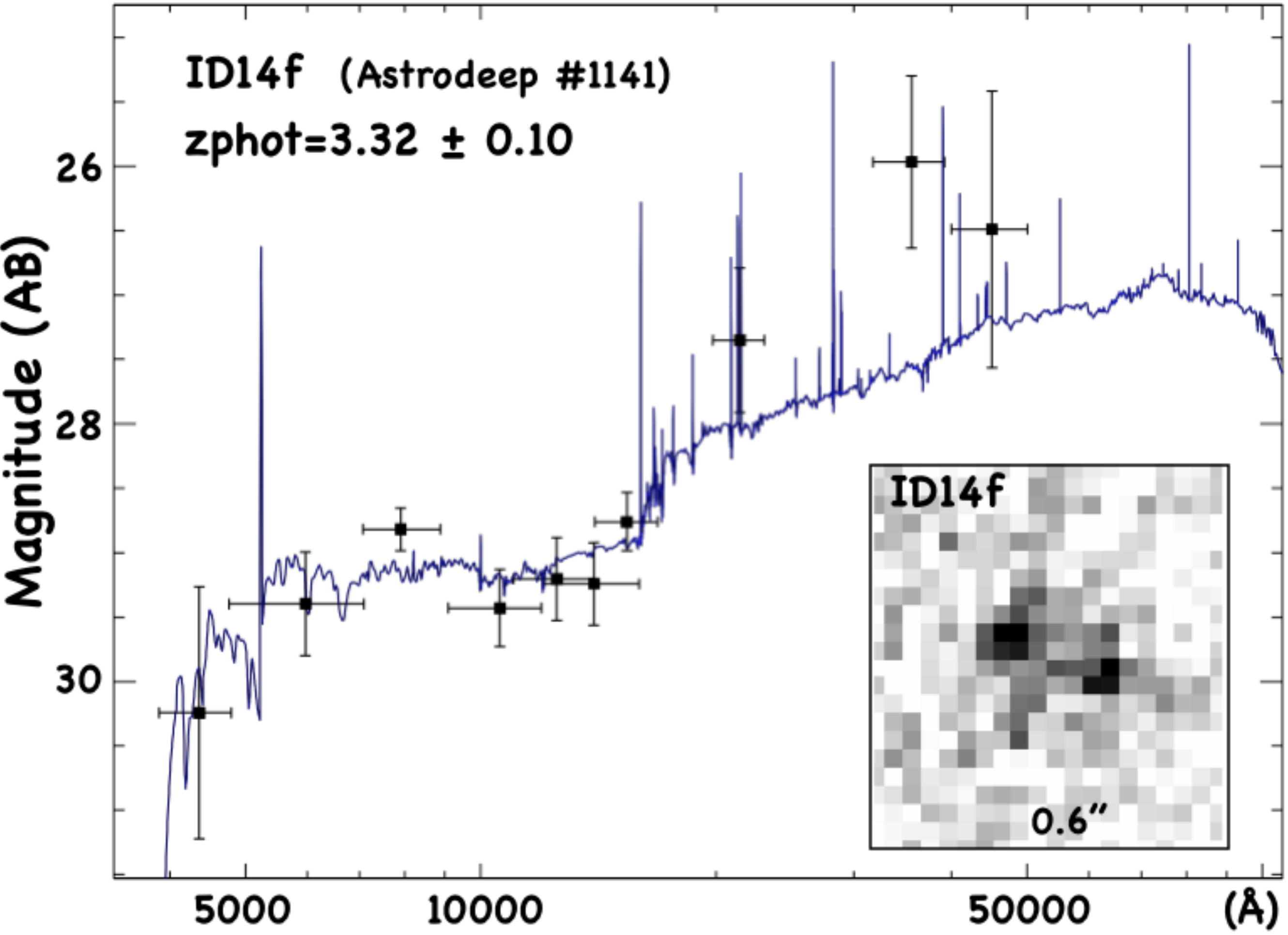} 
\caption{The best-fit SED solution (blue template) compared to the photometry (black points) for image 
ID14f. The photometric redshift with its $1\sigma$ uncertainty is reported (see also \citealt{castellano16b}). 
The inset shows the zoomed ID14f image in the F814W band. The photometric black points includes both components ``1,2''. 
\label{z_phot}}
\end{figure}

\section{ID14, a double lensed system}
The source ID14 is multiply imaged by the cluster MACS~J0416 into three images separated by up to $50''$. The location of these three images is given in  the right panel of Figure~\ref{pano} by the three yellow circles labelled Image1, Image2, and Image3. Image1 happens to be close to a pair of elliptical cluster members, indicated as E1 and E2  in the left panel of Figure~\ref{pano}. These galaxies act as strong lenses,  further splitting Image1 into the four images ID14a, ID14b, ID14c, and ID14d. For consistency, in the following we will refer to Image2 and Image3 as ID14e and ID14f, respectively. 

This system was first discovered by \citet[][]{zitrin13}. In each of the images, the source appears to consist of two bright knots. This morphological marker, together with the image geometry, allowed the first  association of the images to the same source. The two knots are labelled ``1" and ``2" in Figure~\ref{pano}. As discussed in the next section, the images ID14a,b,c,d are spectroscopically confirmed. The association of Image3 with the sixth multiple image of ID14 (namely ID14f) is corroborated by the lens model
recently presented by   \cite{cam16c}, which predicts the presence of an image at this location within $1\sigma$ uncertainty, i.e., $<1''$. 
In addition, the photometric redshift of  ID14f is $zphot=3.32\pm0.10$ (see Figure~\ref{z_phot}),  
which is fully consistent with the spectroscopic redshift of ID14.\footnote{The photometric redshift is also reported
in \citet[][]{castellano16b}. In particular, an example of the best-fit SED for ID14f is shown at following address
{\it http://astrodeep.u-strasbg.fr/ff/?ffid=FF$\_$M0416CL$\&$id=1141}.}

For the analysis discussed in this paper, it is crucial to estimate the magnification of the brightest images of ID14, namely images $a,b$ and $c$. 
 For this goal, the lens model by  \citet[][]{cam16c} could in principle be used. There are, however, two complications. The first is that these images are located in a region of high magnification.  In a collaborative work which involved several groups employing different lens modeling algorithms, and among them the one used by \cite{cam16c}, \cite{meneghetti17} recently showed that the model magnifications in this regime are affected by large uncertainties. For example, at $\mu>10$ the error on the local magnification estimates is $\gtrsim 50\%$. This is due to the strong magnification gradient near the critical lines. The second complication is that images ID14a,b,c are the result of a galaxy-galaxy strong lensing event.    \cite{meneghetti17} also showed that the model uncertainties increase near cluster members.

\begin{figure}
 \epsscale{1.0}
 \plotone{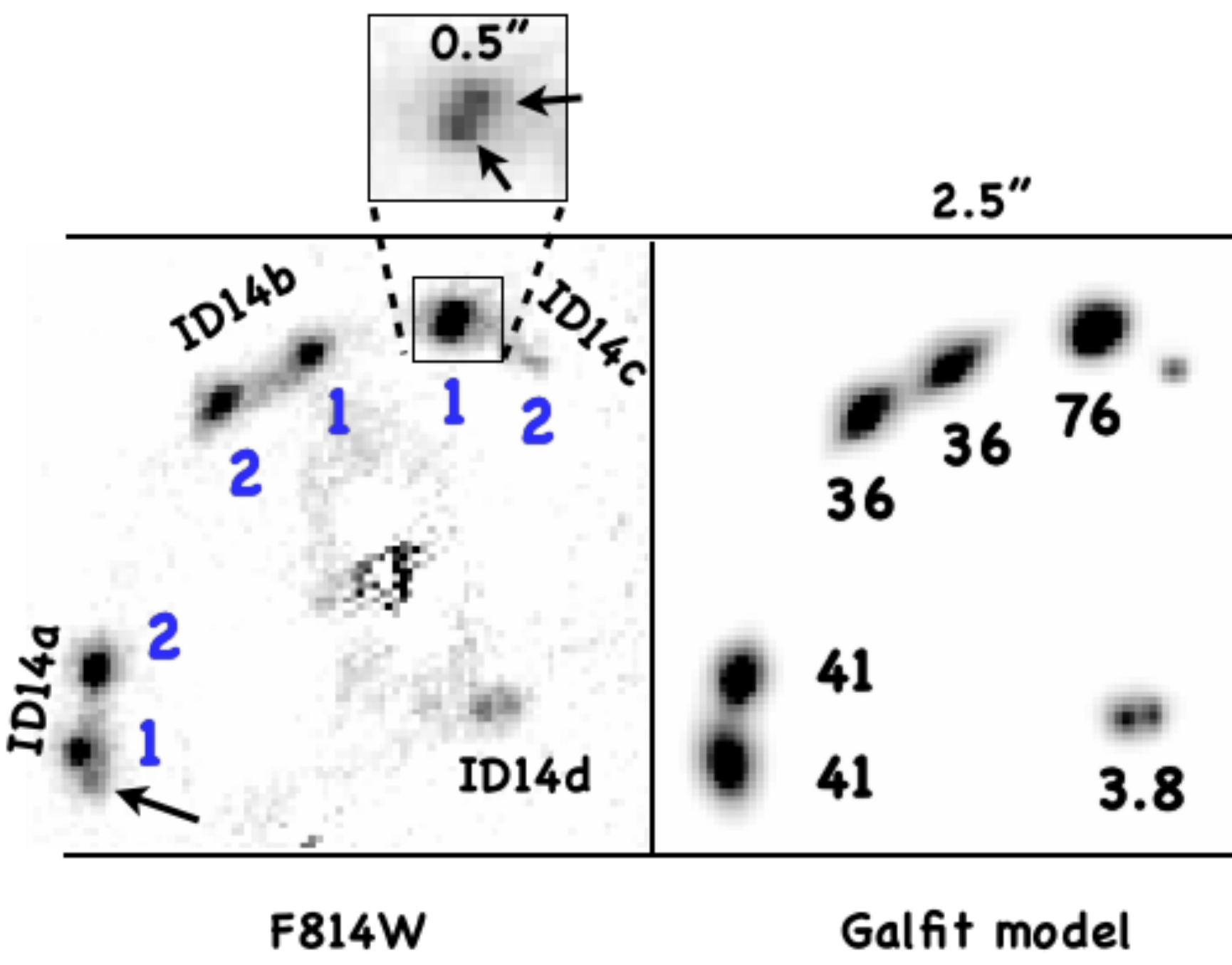} 
\caption{ Left panel: the multiple images ID14a,b,c,d are shown in the F814W band. The morphological details of
images ID14a and ID14c are indicated by the black arrows and the zoomed inset (for ID14c). 
The corresponding {\tt Galfit} models of each component are shown on the right side (the numbers indicate the magnification,
as derived in Sect.~2).
   \label{galfit}}
\end{figure}

For these reasons, instead of reading the magnifications off the map derived from the \cite{cam16c} model, we opted to derive the magnifications of ID14a,b,c using the same method employed in \citet[][]{vanz16a, vanz17b}.  \cite{meneghetti17} showed that magnification estimates are significantly more robust if $\mu <5$. Using the lens model by  \cite{cam16c}, we estimate that the  magnification of ID14f is $\mu=2.0\pm0.1$. This estimate is consistent with others based on different models of MACS~J0416, which we obtained from the Hubble Frontier Field Magnification calculator. If we assume that the model magnification of this image is secure, we can then derive the magnifications of the other images by means of the measured flux ratios.

ID14f is identified with the Astrodeep object \#1411 (ID1411)  \citep{castellano16b,merlin16}. 
It is detected in the F814W, F105W, F125W, F140W and F160W bands, 
with an average magnitude of $29.09\pm 0.15$ resulting from the sum of both components ``1,2''. We computed multi-band, PSF-matched photometry of images ID14a,b,c as described in \citet[][]{merlin16}, but
using the F814W band as detection image instead of F160W. HST photometry has been measured from 30mas
pixel-scale images after subtracting the elliptical galaxies E1+E2 with {\tt Galfit} \citep[][]{peng10}.
T-PHOT v2.0 \citep[][]{merlin16} has been used to measure K and IRAC photometry.
An example of subtracted ellipticals
in the F814W-band is shown in Figure~\ref{pano} (top-left inset), where also the average magnitudes (over the five HST bands) are quoted 
for each component. For example, image ID14a is composed of two components ``1,2'' with magnitudes 26.52 and 26.61,
respectively. The total magnitude, combining the two components, is 25.81. The total magnification for ID14a is therefore 
$\mu$(ID14a)$=\mu$(ID14f)~$\times$~(flux[ID14a]/flux[ID14f]) $= 41\pm7$.
Similarly,  $\mu$(ID14b)$=36\pm7$ and ID14c.1 (the component ``1'' only) is magnified by a factor $76\pm10$.
The errors estimated with this method are $<20$\%, despite the large magnification regime.
The de-lensed magnitude of ID14f(1+2)  is 29.84. Assuming a flux ratio of $\sim 1$ between ``1'' and ``2'' (as observed in images ID14a and b),
the intrinsic magnitude of each component is $\simeq 30.60$.

It is worth comparing the above values with those inferred directly from the Caminha et al. model and other
independent models. The best estimates and the $1\sigma$ statistical errors from Caminha et al. for images ID14a(1) and 
ID14a(2) are $\mu=33_{-12}^{+38}$ and $\mu=23_{-6}^{+15}$, respectively
($\mu=29_{-17}^{+64}$ and $\mu=22_{-11}^{+34}$ for ID14b1 and 2). They change significantly if different and independent
lens models are applied.\footnote{We used the Hubble Frontier Field
Magnification calculator, {\it https://archive.stsci.edu/prepds/frontier/lensmodels/}}
In particular, the magnifications of images ID14a(1) and ID14a(2) inferred from eleven lens models span the interval
 $8 - 110$ with medians $\mu \simeq 30-40$. Also the reported $1\sigma$ uncertainties in each model are
larger than 50\%. The same happens also for the other strongly magnified images, ID14b and ID14c.
The scatter among the various model predictions is of the order of the predicted magnifications,
suggesting that the systematic errors dominate the uncertainties, especially in this complex double-lensed object
(see \citealt{meneghetti17} for a detailed discussion).
Therefore, the method used in this work significantly decreases the uncertainties of the amplification factors
and consequently limits the error budget on the intrinsic physical properties discussed in the next section.

\begin{figure*}
 \epsscale{1.0}
 \plotone{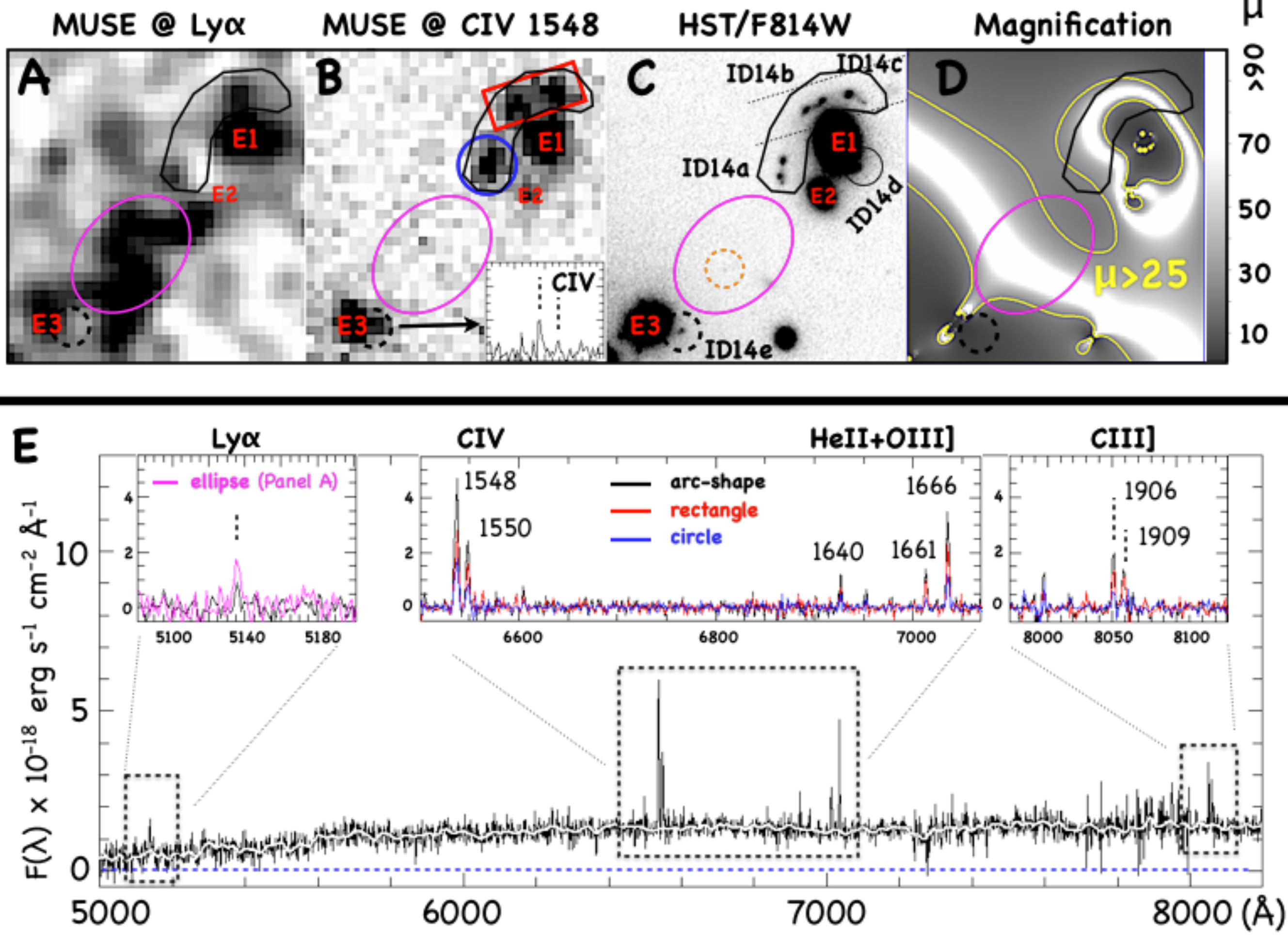}
\caption{Panoramic view of the MUSE spectra. Top panels: the zoomed region centered on ID14a,b,c,d,e is shown ($6.8''$ wide). 
From left to right we show the MUSE images at the \lya\ (panel A), at 1548 \AA~(panel B), in the HST/F814W band (panel C), 
and the gray-coded magnification map (panel D), in which the yellow contours enclose the region with $\mu>25$.
The arc-shaped black line marks the aperture used to extract the MUSE spectrum as the sum of images ID14a,b,c.
The black dashed circle identifies the image ID14e and the magenta ellipse the diffuse \lya\ emission at the same
redshift as ID14 on top of the critical line of the galaxy cluster. Panel (B) shows the rectangular red aperture with the same
width and position angle of the X-Shooter slit (also shown in panel C, top-right, with black dotted thin lines).
The orange dotted circle in panel C marks the position of a possible counterpart of the diffuse \lya\ emission
(magenta ellipse). 
The blue circle ($1.2''$ diameter) marks the aperture used to derive the spectrum for image ID14a only. 
Panel (E), shows the MUSE spectrum of ID14a,b,c (black line) extracted using the arc-shaped
aperture. The continuum of the elliptical galaxies E1+E2 has been subtracted (white thick line) and the high ionization lines
are shown in the corresponding insets. The colors of the spectra in the insets corresponds to the colors of the apertures
shown in panel B. The flux scale of all images is $F(\lambda)$ [$10^{-18}$ \erg].  \label{muse}}
\end{figure*}

%
\begin{deluxetable}{llll}
\tabletypesize{\scriptsize}
\tablecaption{Observed spectral lines. \label{infos}}
\tablewidth{0pt}
\tablehead{
\colhead{line/$\lambda_{vacuum}$} & \colhead{[$F_{obs}$]($\frac{S}{N}$)($\sigma_v$)} & \colhead{[$F_{corr}$](EW)} & \colhead{Redshift}}
\startdata
\lya\ 1215.69                                                                         &    [0.24](2.5)(--)      &  [0.09]($\simeq1.1$) &   [M] (3.226)\\     
{[\textrm{C}\textsc{iv}]\ensuremath{\lambda1548.20}}  &    [2.18](20.0)($\lesssim64$)         &  [0.87](16.9)              & [M] 3.222 \\ 
{[\textrm{C}\textsc{iv}]\ensuremath{\lambda1550.78}}  &   [1.20](11.0)($\lesssim64$)          &   [0.48](9.3)              & [M] 3.223 \\ 
{\textrm{He}\textsc{ii}\ensuremath{\lambda1640.42}}  &    [0.33](4.8)($<38$)     &    [0.13](2.8)            & [M] 3.223 \\ 
{\textrm{O}\textsc{iii}]\ensuremath{\lambda1660.81}}  &    [0.48](5.4)($<42$)     &    [0.19](4.3)            & [M] 3.222\\  
{\textrm{O}\textsc{iii}]\ensuremath{\lambda1666.15}}  &     [1.02](11.7)($<42$)  &   [0.41](9.2)          & [M] 3.222\\  
{[\textrm{C}\textsc{iii}]\ensuremath{\lambda1906.68}} &     [0.55](6.2)($<42$)    &   [0.22](6.5)         & [M] 3.223\\  
{\textrm{C}\textsc{iii}]\ensuremath{\lambda1908.73}}  &     [0.45](5.1)($<42$)    &   [0.18](5.3)         & [M] 3.222\\  
{[\textrm{O}\textsc{ii}]\ensuremath{\lambda3727.09}}  &     [$<0.28$](1)               &  --                       &  [X] --\\
{[\textrm{O}\textsc{ii}]\ensuremath{\lambda3729.88}}  &     [$<0.28$](1)               &  --                       &  [X] --\\
{[\textrm{O}\textsc{ii}]\ensuremath{\lambda4364.44}}  &     [$<0.28$](1)               &  --                       &  [X] -- \\
$H\beta \lambda 4862.69$                                                &    [1.06](3.7)(--)                &    [0.71](136)           &   [X] (3.2222)\\   
{[\textrm{O}\textsc{iii}]\ensuremath{\lambda4960.30}}  &    [2.76](9.6)($<23$)       &    [1.84](367)           &   [X] 3.2222\\     
{[\textrm{O}\textsc{iii}]\ensuremath{\lambda5008.24}}  &    [8.92](31.5)($<18$)     &    [5.95](1209)           &   [X] 3.2222\\    
\hline
\lya\ 1215.69                                                                         &    [0.85](7.4)($<40$)       &     --     &   [M] 3.223\\    
\tableline
\tablecomments{
Column \#1: the observed emission lines and their 
rest-frame wavelengths. Column \#2: observed fluxes 
(in units of $10^{-17}$\ergscm), S/N and $\sigma_{v}$ (instrument-corrected velocity dispersion in \kms)
for ID14 extracted from the apertures that include the multiple images.
Flux limits are reported at $1\sigma$.
Column \#3: Fluxes associated to the object ID14 (components ``1,2'') used to derive flux ratios and intrinsic quantities 
($F_{corr})$ (see text for details). The rest-frame
EW (\AA), derived adopting the SED-fitted continuum of ``1''+``2'' ($m\simeq25.80$)
is also reported. The de-lensed fluxes can be obtained by dividing these values by $\mu=40$.
In column \#4 the MUSE and X-Shooter redshifts are indicated
with [M] and [X], respectively (redshifts in parenthesis are uncertain). 
The $\sigma_{v}$ of the lines with $S/N>3$ are estimated from the higher resolution
X-Shooter spectrum when possible, or from MUSE otherwise.
The \lya\ flux calculated within the magenta ellipse (whose area is 6.7 sq. arcsec) shown in 
Figure~\ref{muse} is  reported in the last row.}
\end{deluxetable}

\section{Characterizing system ID14}
The main goal of our investigation is to estimate the size and the age of our system, as 
well as to assess the nature of the radiation field originated by its two close, compact sources.   
The gas phase metallicity is also estimated, as well as the dynamical mass, useful to 
constrain the baryon content of ID14. 

\subsection{Sizes of the components ``1'' and ``2''}

{\tt Galfit} fitting \citep[][]{peng10} has been performed
on both components  ``1,2'' of images ID14a,b,c  (following the methodology described in  \citealt[][]{vanz17b})
after the elliptical galaxies, E1 and E2, have been subtracted (Figure~\ref{pano}, top-left). 
All the multiple images are very nucleated and tangentially elongated along the arc shape
showing an average effective radius $<R_e > =1.5\pm0.3$ pix and an axis ratio ($q=b/a$) $<q>=0.3\pm0.2$
(an example of {\tt Galfit} modeling is shown in Figure~\ref{galfit}).
The circularized radius $R_{c}$ ($=R_{e}\times q^{0.5}$) is 0.8 pix (1 pix = $0.03''$) corresponding 
to a de-lensed size of $\simeq30$ pc at z=3.222, adopting $\mu=40$, 
(or sizes lower than 50 pc if we increase $R_e$ to 2 pix and decrease $\mu$ to 30).
The two components ``1,2'' are separated by 300 pc on the source plane.

Additional details on the observed multiple images are highlighted in Figure~\ref{galfit} (see ID14c), 
but we will not discuss them further in this work. 
We stress that these additional features are at least two magnitudes fainter than the main components 
(or constitute sub-structures of the components themselves) and, given the large magnification, they would have a de-lensed
ultraviolet magnitude fainter than $31$ and de-lensed sizes possibly consistent with forming star-clusters on parsec-scale.
These additional features might also be the result of the extremely large magnification gradient characterizing this field.

\subsection{Spectral features}

\begin{figure*}
 \epsscale{1.0}
 \plotone{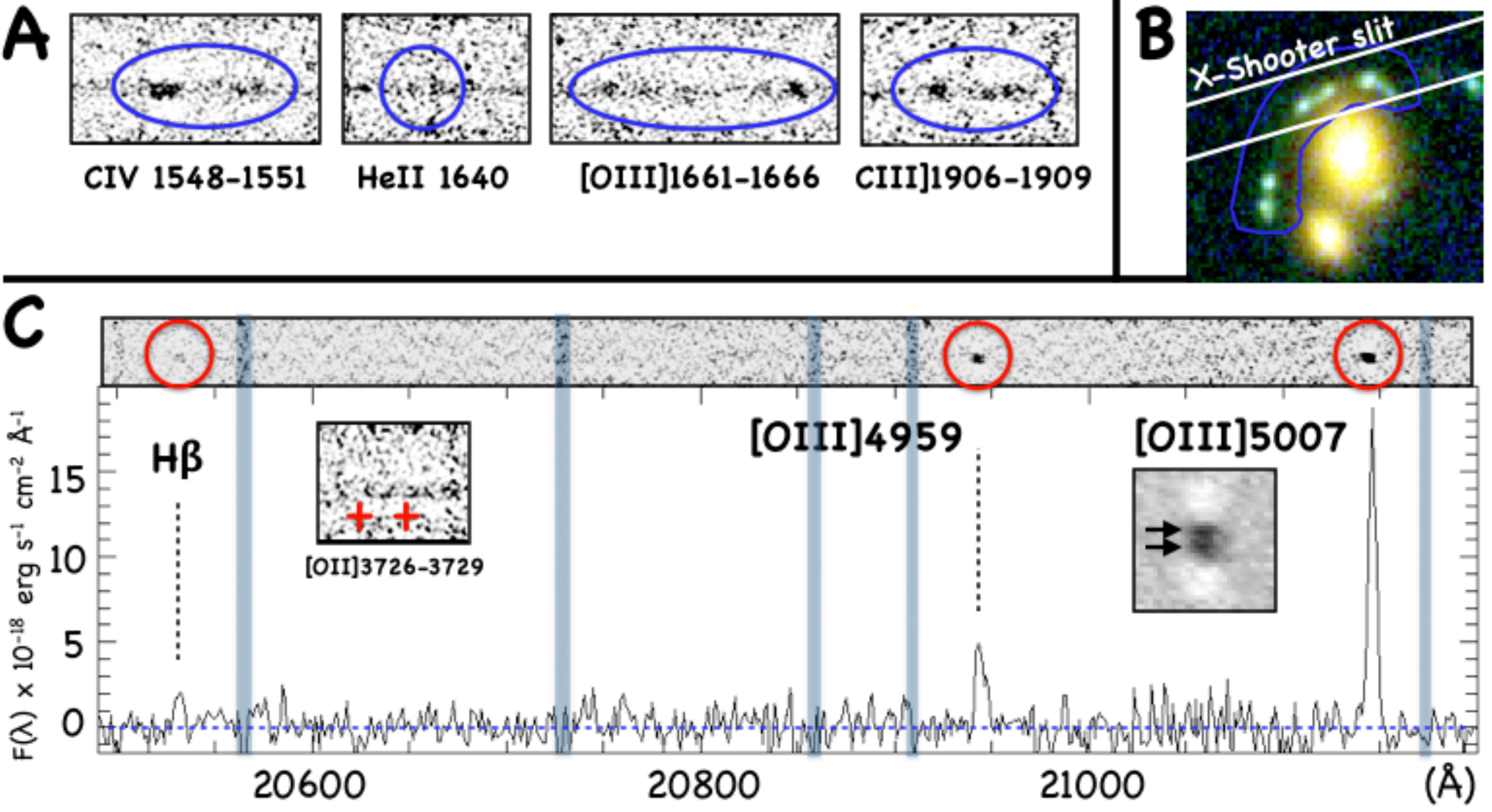}
\caption{The X-Shooter spectra in the VIS and NIR arms are shown for the relevant emission lines. In panel (A)  the two-dimensional
zoomed spectra of the UV metal high-ionization lines are shown (the faint continuum comes from the foreground elliptical galaxy). 
In panel (B) the color image of ID14a,b,c,d is shown with the MUSE aperture used to extract the spectra
(blue curve) and the orientation and width ($0.9''$) of the X-Shooter slit (white lines). The size of the image is $3''$ on a side.
The one and two-dimensional \hb\ and \oiiidoub\ lines observed at $\simeq 2 \mu m$ are shown in panel (C). 
The left inset shows the two-dimensional spectrum centered at the position of the
expected \oiidoub\ doublet. The right inset is a zoomed image of \oiiiv, in which ID14b and ID14c 
are spatially resolved (the arrows mark their position).  \label{xsho}}
\end{figure*}

The 2h VLT/MUSE (ID 094.A-0115B, PI: J. Richard) and 2h VLT/X-Shooter  (ID 098.A-0665B, PI: E. Vanzella) 
spectra of images ID14 are shown in Figures~\ref{muse} and \ref{xsho},
obtained with good seeing conditions, $0.7''$ and $0.5''$, respectively.
The apertures over which the spectra have been extracted are also shown in the same figure. Data reduction for X-Shooter and
MUSE have been performed as described in \citet[][]{vanz16a} and \citet[][]{cam16c}, respectively. 

\begin{figure}
 \epsscale{1.0}
 \plotone{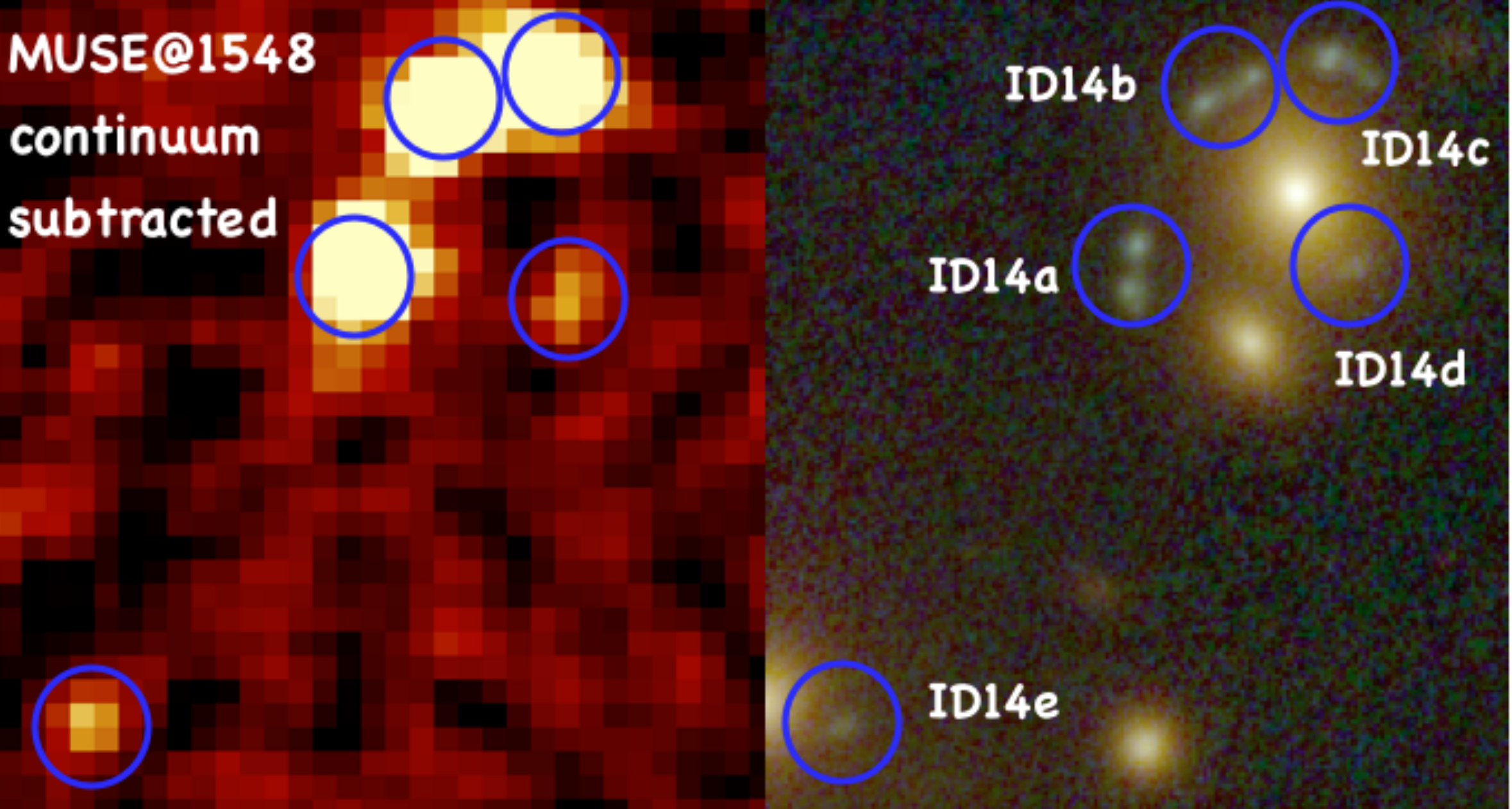}
 \caption{The continuum-subtracted data-cube at the \civblue\ wavelength position is shown on
the left panel ($6.5''$ wide). The blue circles mark the multiple images
of ID14 detected on the HST color image, reported in the right panel. In particular,
faint line emissions arise at the position of multiple images ID14d and ID14e. \label{CIV}}
\end{figure}

The most prominent line in the MUSE spectrum is the \civblue\ component of the \civalone\ doublet. We use this line to 
measure the
redshift of the multiple images contained in the data-cube (only ID14f is not included in the MUSE field of view).
A continuum-subtracted data-cube has been computed by averaging the signal in a central window of 3 spectral elements 
(one element corresponds to 1.25\AA) and subtracting the average signal from two adjacent 
windows of 10 spectral elements each \citep[see also][]{vanz17a}. Figure~\ref{CIV} (left panel) 
shows the result. The  \civblue\ line is detected for  all images ID14a,b,c,d,e with S/N$>5$. 

Thanks to lensing  magnification, the photometric properties
and sizes of  components ``1'' and ``2'' can be studied individually. On the other hand, the ultraviolet and optical spectra
are seeing-limited and contain the contribution of both  components.
It is worth noting that the \civalone\ emission (especially for ID14a) shows an elongated shape
that follows the spatial orientation of components ``1,2''. This suggests that both components contribute to the
line emission and in general to the ultraviolet metal lines. 
Possibly, MUSE equipped with the adaptive optics module may allow these spectral
features to be spatially resolved. 

We exploited the MUSE data-cube by choosing a free-form aperture encompassing the three images ID14a,b,c 
(increasing the S/N); on the other hand, the X-Shooter slit captures the image ID14b and one component (``1'') of the image ID14c 
(being the component ``2'' less magnified, see Figure~\ref{xsho}, top left). 
Line fluxes, signal-to-noise, velocity dispersions, rest-frame equivalent widths (EW) and redshifts are reported in Table~\ref{infos}. 
High ionization lines \civ, \heii, \oiiiuv\ and \ciiidoub\ have been detected with S/N ratios ranging between 2 and 20 and the
X-Shooter lines, \hb, \oiiidoub, with S/N=$4-31$. High ionization lines have also been confirmed with X-Shooter (with $R=7400$),
though only \civblue\ and \oiiiuvred\  have S/N$\gtrsim3$ (see Figure~\ref{xsho}). 
Remarkably, no \lya\ emission has been detected (S/N $<3$) in the same spatial region where both the metal lines and
stellar continuum arise. We discuss this in Sect.~4.3. 
The observed line fluxes derived from the aforementioned apertures are reported in the column \#2 of Table~\ref{infos}. 
Since they include different
portions of the multiple images, we rescaled properly the fluxes in column \#2 to obtain the values associated to the single object 
ID14 (``1,2''). The corrected fluxes are reported in column \#3, $F_{corr}$, 
obtained by dividing the measured fluxes reported in column \#2 by 2.5 and 1.5 in the case of MUSE
 and X-Shooter observations, respectively. The two rescaling factors have been derived from specific apertures
 defined in the MUSE data-cube:  a circular aperture centered on ID14a and a rectangular aperture mimicking the X-Shooter slit 
(see Figure~\ref{muse}, panel B). 
The equivalent widths of the lines have been derived using $F_{corr}$ and the inferred continuum magnitude
from the SED-fitting of both components ``1,2'' (see Sect. 3.3). Typical  errors mainly reflect the uncertainty in the
line flux, as the underlying images are well detected in the deep HFF images.  Conservatively, even including the possible
systematic errors (due to the subtraction of E1+E2) the final error on the EWs is not larger than 50\%. 

Beside the presence of ultraviolet metal lines, also the prominent \oiiidoub\ line emission, the
ratios $\oiiidoub / \oiidoub\ > 15$ and \oiiidoub / \hb\ $>10$ suggest a highly ionized medium and a possibly
large \lya\ escape fraction \citep[][]{erb16, henry15}. 
Despite that, the \lya\ emission in ID14 is deficient (see Sect. 4), independently from the adopted aperture shape.

\subsection{Physical Properties}

SED fitting has been performed using BC03 templates \citep[][]{BC03} on each component ``1,2'', for  images 
ID14a,b and the component ``1'' of ID14c (see \citealt[][]{castellano16b})
after subtracting the elliptical galaxies E1+E2 in all the HST images and using the T-PHOT
v2.0 algorithm, \citet[][]{merlin16} (see Sect.~2).
The subtraction of E1+E2, however, may introduce not negligible color trends,
especially in the near-infrared bands. Such residuals may introduce systematics 
in the inferred physical quantities.
In order to add constraints to our photometric analysis, we also estimate the dynamical mass and the gas phase metallicity.
The inferred dynamical mass from the
$\sigma_v$(\oiiiv) $\lesssim 20$ \kms\  velocity dispersion is $M_{dyn} < 2 \times 10^{7}$ \msun,
where we assume the \oiiiv\ line width is associated to both components ``1,2'' of 30pc
radius each \citep[see discussion in][]{vanz17b, rhoads14, erb14}. 
The metallicity is computed from the high-ionization and optical rest-frame line ratios \citep[][]{perez17, pilyugin06}.
The electron temperature $T_e = 19000\pm1000$K is derived from the
observed \oiiiuv\ / \oiiiv\  ratio. 
our estimate of $T_e$ is not sensitive on the assumed electron density in the range 
$n_e=10-10000$ cm$^{-3}$. We obtain 
12+log(O/H)~$\simeq 7.7\pm0.2$ that corresponds to about 1/10 solar.
We performed the SED-fitting by fixing the metallicity to the measured value and considering only those 
solutions producing stellar masses not exceeding 10 times the estimated
dynamical mass (the factor ten includes the uncertainties in the $M_{dyn}$ estimate).
In general, the inferred quantities agree among the multiple images of the same component.

\begin{figure*}
 \epsscale{1.0}
 \plotone{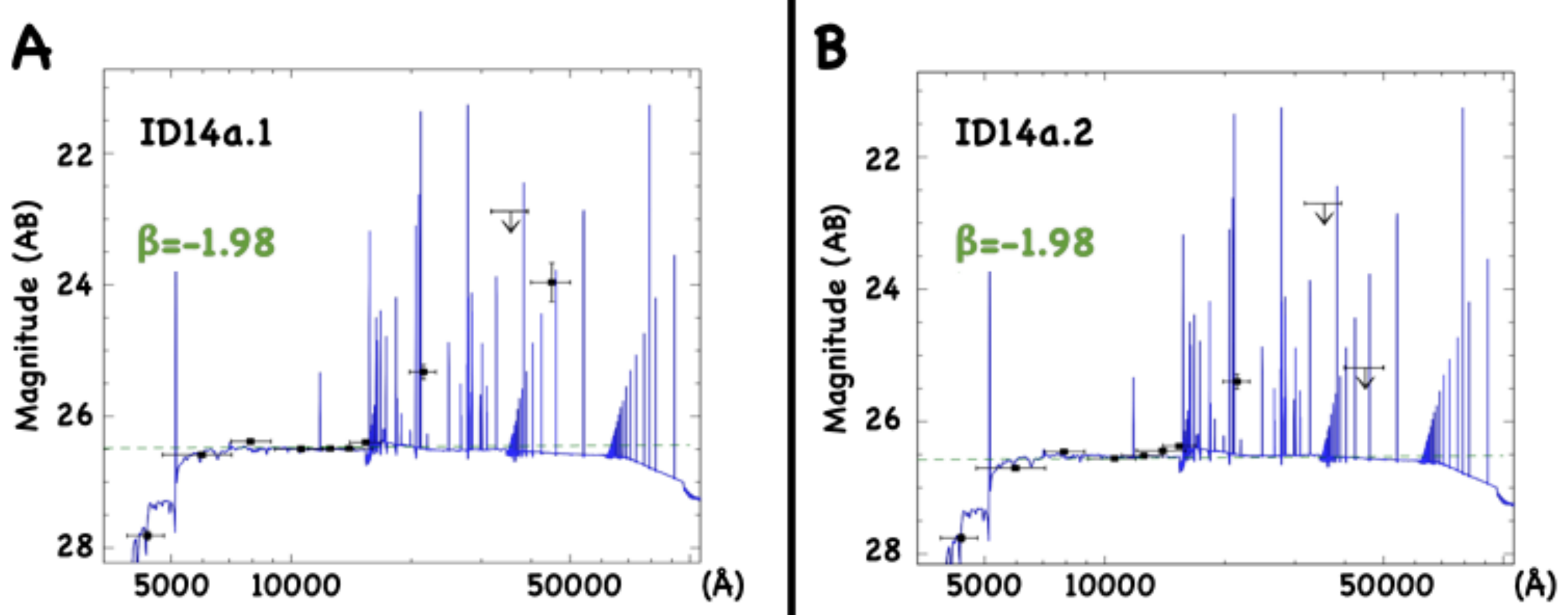} 
\caption{Best-fit SED solutions for the components 1 and 2 of the ID14a image. The inferred 
physical quantities are reported in Table~\ref{infos2}. 
The greed dashed line is the ultraviolet slope $\beta$. The photometric discontinuity at 22000 $\AA$  
corresponds to the K-band and is fully consistent with the measured flux of the optical emission lines observed 
with X-Shooter (see text). 
   \label{SEDs}}
\end{figure*}

In particular we focus on images ID14a.1 and ID14a.2 that are the least contaminated
by the elliptical galaxies E1+E2. The best-fit yields 
stellar masses $\lesssim 10^{7}$\msun, ages $\lesssim 100$ Myr, SFRs$\simeq$0.1 \msunyr\  and E(B-V)$<$0.1.
Table~\ref{infos2} summarizes the results and Figure~\ref{SEDs} shows the best-fit solutions.
It is worth noting the $\simeq 1$ magnitude excess in the K-band is fully consistent with the observed
\oiiidoub\ flux. In particular the measured equivalent width of EW(\oiiiv)$\simeq$1200\AA\ rest-frame, 
the inferred metallicity, young ages  and low stellar mass place this
object in the category of the so-called ``Green Pea'' (GP) sources, though the
stellar mass and size are in our case an order of magnitude lower than those typically reported
\citep[][]{cardamone09, amorin15, amorin17, henry15}. 
In particular, \citet[][]{henry15} selected a sample of 10 Green Pea sources ($z\sim0.2$) 
on the basis of high equivalent width optical emission lines and found strong \lya\ emission in all of them. 
Remarkably, despite ID14 shows many properties in common with the GPs sources,
the \lya\ is deficient (see next section).

\begin{deluxetable}{lccc}
\tabletypesize{\scriptsize}
\tablecaption{Properties of Images ID14a and ID14f. \label{infos2}}
\tablehead{
\colhead{~} & \colhead{ID14a(1)} & \colhead{ID14a(2)} & \colhead{ID14f(1+2)}}
\startdata
M$_{stellar}$ [$10^{6}$\msun]  & $2.5-9.8$  &  $2.0-7.6$  & $-$  \\     
SFR [\msunyr]              &  $0.10 - 0.15$  & $0.10-0.17$    &  $-$\\ 
Age [Myr]                      &   $16-100$     &  $13-80$          &  $-$\\ 
E(B-V)                           &   $\simeq 0.06$ & $\simeq 0.06$       &$-$ \\ 
$R_{c}$ (UV) [pc]        &   $30\pm11$             &  $30\pm11$           & $-$\\  
$m_{UV}$(obs)  &  $26.52\pm0.03$    &  $26.61\pm0.03$      & $29.09\pm0.15$\\  
$m_{UV}$(int)    &  $30.55\pm0.19$         &  $30.64\pm0.19$          & $29.84\pm0.16$ \\  
M(1500\AA)(int)           &  -15.12                  & -15.03                     &  -15.85  \\
$\beta_{UV}$               &  $-1.98\pm0.20$            & $-1.98\pm0.20$        &   $-2.0\pm0.5$ \\
$\mu$                            &   $40\pm7$               &  $41\pm7$                 &    $2.0\pm0.1$\\
\hline
\tableline
\tablecomments{
 De-lensed physical properties are derived from SED-fitting
for the single components ``1'' and ``2'' of image ID14a and the combination ``1+2''
of ID14f by adopting the measured metallicity (see text).
The 68\% central intervals are reported for the  stellar mass, star formation rate and age,
and the 1$\sigma$ errors for the remaining quantities. 
The observed (``obs'') and intrinsic (``int'') magnitudes are also reported.
The intrinsic stellar masses and SFRs are obtained dividing by $\mu$
the best fit parameters resulting from the SED fit.
It is worth noting that the values for  ID14f include both components ``1+2''.
The magnification $\mu$ is also reported and derived as described in the text. 
}
\end{deluxetable}

\section{Discussion}

\subsection{The nature of the ionizing source}
\begin{figure}
 \epsscale{1.0}
 \plotone{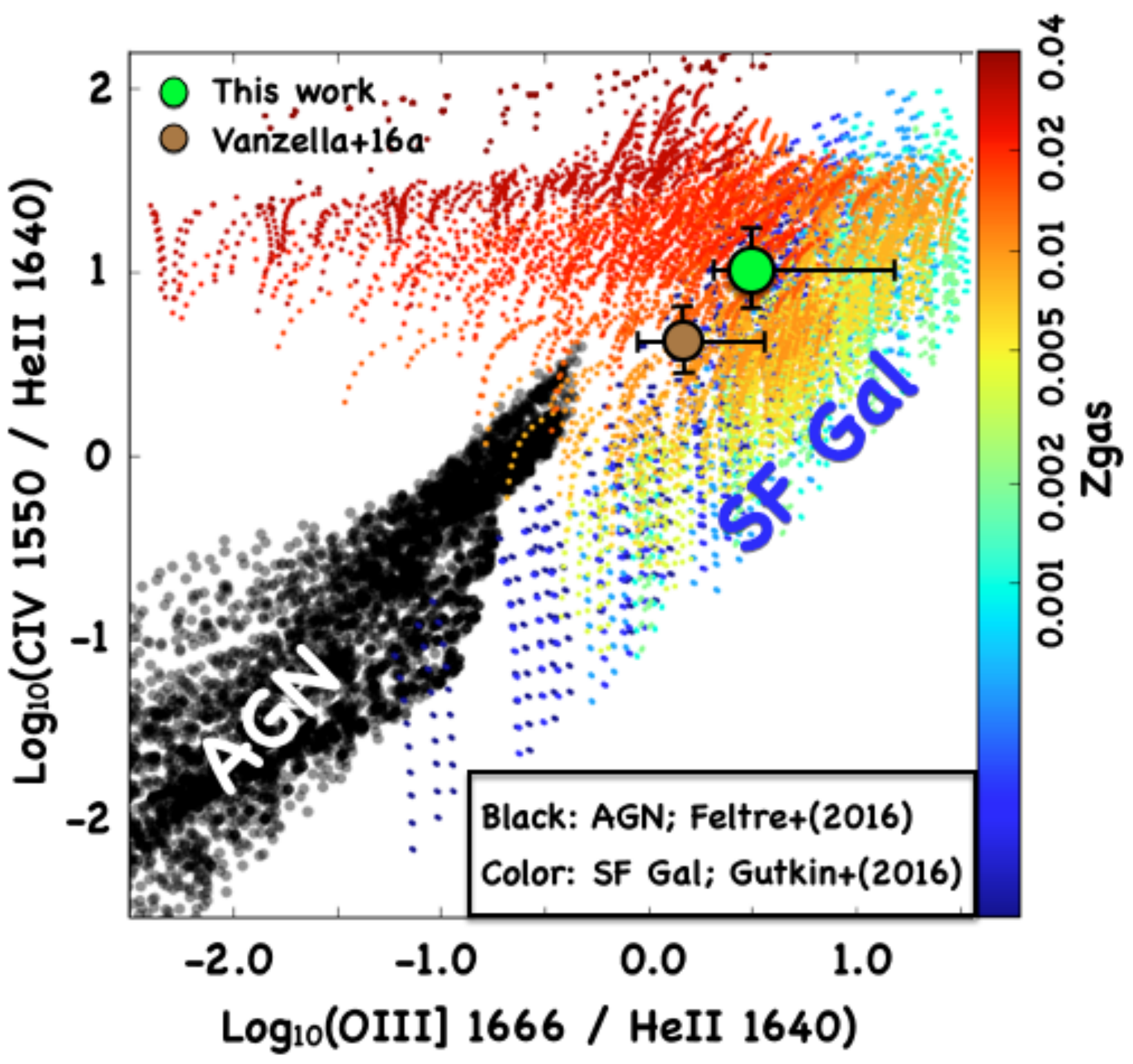} 
\caption{A diagnostic diagram separating AGN and star-forming powered sources
with superimposed object ID14 and a similar object described in \citet{vanz16a}. 
Predictions are from photoionization models of \citet[][]{feltre16} (AGNs, gray filled circles) and \citet[][]{gutkin16}. 
Star-forming galaxies are color-coded according to their gas phase metallicity, $Z_{gas}$, as defined in \citet[][]{gutkin16}.
   \label{diagrams}}
\end{figure}

ID14 shows similarities with other star-forming systems identified at $z=3-7$ \citep[][]{stark15,vanz16a,mainali17,schmidt17}.
We used the large grid of photoionization models described in \citet[][]{feltre16} and \citet[][]{gutkin16} to investigate
the nature of the ionizing source. 
In all the diagnostic diagrams, ID14 belongs to the regions powered by star-formation (e.g., see Figure~\ref{diagrams}).
The main reason for that is the faintness of the \heii\ line when compared to \civ\ and \oiiiuv, that would suggest a drop of the spectrum
at energies just above 4Ry \citep[see][]{mainali17}. 
From the archival 300~ks Chandra exposure \citep{ogrean15}, we derived 2$\sigma$ upper limits on the X-ray flux 
of $1.6 \times 10^{-16}$ and $8.3\times 10^{-16}$ \ergscm\ in the soft (0.5-2 keV) and hard (2-7 keV) band, respectively,
taking into account the strong background emission due to the intra-cluster gas. 
At the redshift of ID14 these limits correspond to
luminosities of $1.3\times10^{43}$ and $6.1\times10^{43}$ \lumi\ in the soft and hard band, respectively.
They cannot exclude the possibility that an absorbed Seyfert-like AGN could be present. 

In addition to the above limits and the line ratios, the fact that components ``1,2'' are spatially resolved and show very narrow
nebular emission lines further supports the evidence that the source of ionizing radiation is dominated
by stellar emission. This is also reminiscent of
recent studies in the local universe \citep[e.g.,][]{smith16,annibali15,kehrig15}, where young super star clusters show narrow
high ionization nebular lines. Although a low-luminosity AGN can not completely excluded, the presence of sub-structures in
ID14 and the aforementioned characteristics support the conclusion that the star-formation is likely to be the main source
of ionizing radiation. 

\subsection{A L=0.004L$^{\star}$ super star clusters at z=3.2222}

ID14 is made of two main star-forming regions (``1,2'') of $\simeq 30$ pc effective radius each, with de-lensed 
magnitudes $\simeq 30.6$ (L=0.004L$^{\star}_{z=3}$), possibly showing additional fainter sub-components. 
The low stellar mass ($<10^{7}$\msun), low metallicity, young ages ($\lesssim100$ Myr)
and hot stars are also distinctive features of this system.

The double knot morphology of ID14 suggests two super star clusters of several $10^{6}$~\msun\  are present and 
possibly composed by a mixture of stellar populations. In fact, the high-ionization (e.g., \heii\ and \civ) lines require 
a young  ($\lesssim10$~Myr) stellar component, characterized by blackbody effective temperatures higher
than 50000~K \citep[][]{steidel14,stark15}, and the ages inferred from the SED fitting also support the presence
of an underlying older ($\lesssim100$~Myr) population.

The inferred specific star formation rate (sSFR) is large ($>20$ Gyr$^{-1}$), a value comparable to those
inferred by \citet[][]{stark14} and \citet[][]{karman17}  and suggesting the system is still in a starburst phase and rapidly growing. 
Conservatively, adopting the lowest stellar mass ($2.5\times10^{6}$~\msun) and the largest effective radius ($R_{C}=50$ pc) 
of the 68\% intervals derived above, the observed stellar mass density of the star-clusters 
is relatively large, $>150$\msun$pc^{-2}$ \citep[see also,][]{vanz17b}. 

\subsection{Deficient  \lya\ emission in ID14}

From the \hb\ line flux,  an estimate of the theoretical \lya\ emission
can be computed as f(\lya) = $8.7 \times f(\ha) \times 10^{1.048\times E(B-V)}$, where we assume a case B
recombination theory, f(\ha)/f(\hb)=2.7 \citep[][]{brocklehurst71}.   
Adopting E(B-V)=0.0(0.2) (for the nebular emission), the expected intrinsic \lya\ flux is
$1.7(2.7)\times10^{-16}$\ergs,
i.e., more than 150 times brighter than what was observed, and more than 12 times brighter than \civ. 
Despite of that, the observed ratio (\civ / \lya) is $\gtrsim 15$. Therefore, \lya\ emission is strongly deficient. 
A conversion of the intrinsic \lya\ emission into absorption can in principle be
achieved by dust suppression of \lya\ photons,
by geometrical effects leading to the scattering of \lya\ photons out of the
line of sight, or by a combination of both. 
Similar physical conditions observed in local low-metallicity star-forming dwarf galaxies
might help to better understand the nature of ID14 (though at different scales). For example, the low-metallicity
star-forming dwarf galaxy IZw~18 shows \lya\ absorption, despite its low
metallicity and young age. Also, galaxies of the LARS sample (or: local, \ha\
selected galaxies) with low velocity dispersion show a reduced \lya\ escape \citep[][]{herenz16}.
A large number of scatterings in static, high column density gas can lead to
an efficient suppression of \lya\ photons by even small amounts
of dust. The \lya\ absorption could also be caused by diffusion of the
photons out of the observer direction (not by dust destruction).
As a result, \lya\ absorption can be observed also in dust-free galaxies
\citep[see, ][]{atek09}.
\citet[][]{verhamme12} show directional variation in the \lya\ escape from a
simulated disk galaxy (see also \citealt{zheng14}). This directional variation is reduced significantly
in the presence of outflows \citep[e.g.,][]{duval16} and/or for non-disk
geometries including modified shell models \citep[][]{beh14} and/or clumpy ISM
models \citep[][]{gronke14}, where the \lya\ escape fraction varies by only a
factor of order unity
and thus do not provide the variations required to explain the
observed flux ratio. However, the theoretical studies have considered
systems with $\sim$kpc extents, and ignored observational aperture
effects. The anisotropic escape of \lya\ is strongest if the source is
``shielded'' by a gas cloud of similar or bigger size than the emitting
region (in this case few tens parsec), 
and weakened by photons scattering back into the
line-of-sight. Therefore, having a very compact source as in this
work is likely to enhance the directional variations, but  
further studies are needed to explore this possibility. Observational 
aperture effects are not an issue in the present case, given the wide MUSE FoV.
The strong lensing effect can in principle modulate the spatial appearance of the \lya\ emission, 
however in the present case no \lya\ emission is observed on top of the stellar ultraviolet continuum
of images ID14a,b and ID14c.
Plausibly, ID14 would appear as a \lya-emitter if observed from
a different view-angle and would require a variation of the \lya\ intensity of
two orders of magnitudes. 
A possibility is that even a small amount of dust extinction
in a region enshrouded by neutral gas (or a partially neutral IGM) 
may explain the strong depression of the \lya\ line.
It is also worth noting that 
the \civ\ position measured with X-Shooter does not show clear signatures of outflowing gas
(at velocities $>50$\kms, see also \citealt[][]{vanz16a}), further supporting an
efficient \lya\ attenuation.  

\begin{figure*}
 \epsscale{1.0}
 \plotone{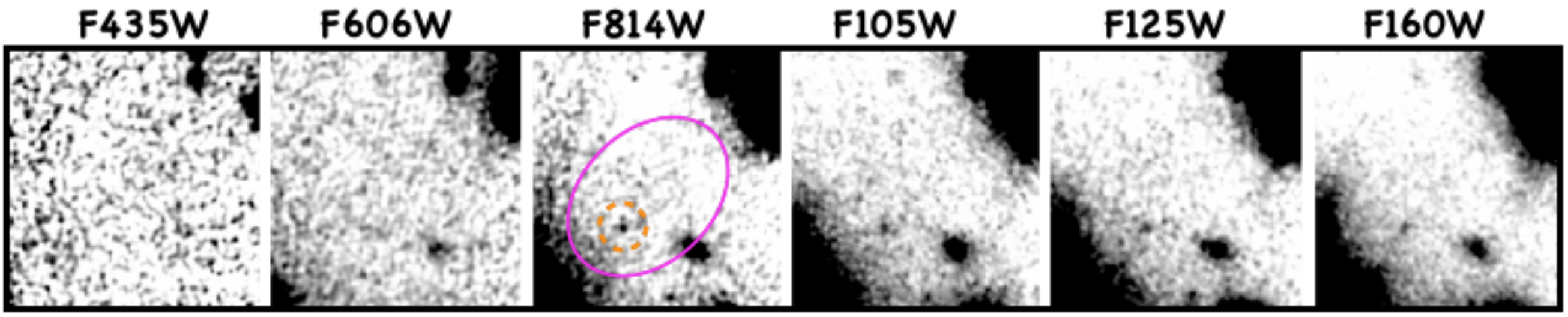} 
\caption{HST/ACS and WFC3 cutouts ($1.5''$ wide) centered on the diffuse \lya\ emission marked with the magenta ellipse. 
 The dashed-orange circle marks the position of the faint $m \sim 30$ (or $m>34$ de-lensed) underlying object if placed at the same redshift of the
 \lya\ emission.
   \label{orange}}
\end{figure*}

\subsection{A spatially offset ultra-faint \lya\ emission}

As discussed in the previous section, ID14 is a system with prominent high-ionization lines with
energetic photons reaching at least 4 Ry. The deficiency of the \lya\ emission at the spatial position where the 
metal lines and the stellar continuum arise suggests an efficient attenuation by dust and/or gas. 
However, such process could be not so efficient in other directions. 
In the case of a non-uniform covering factor of the neutral gas  
we may expect to recover the diffused \lya\ photons in other directions and/or observe spatially-offset 
fluorescent \lya\ emission \citep[][]{furlanetto05, weidinger05}. 
Indeed, we detected an offset and spatially-extended \lya\ emission (S/N=7.4) at the {\it same} redshift as ID14
(separated by $\simeq 2.1$ kpc) that may corroborate this possibility. The observed 
\lya\ emission lies  across the critical line produced by the galaxy cluster at the given redshift. The position of the
critical line (the highest magnification) is strongly
supported by the confirmed images of ID14, ``image1'' and ``image2''  (Figure~\ref{pano}, panel B). 
This also means the \lya\ is strongly magnified and consequently highly spatially distorted (see Figure~\ref{muse}, magenta ellipse).
The observed emitting region (6.7 arcsec$^2$) corresponds to de-lensed 7.6 kpc$^2$ 
and a de-lensed integrated flux of $1.7\times10^{-19}$\ergscm\ if we adopt
a median magnification $\mu=50$ (calculated within the same region).
 However the real physical size and shape 
of the emission is affected by large uncertainties, especially where the magnifications 
formally diverge.
This spatially offset \lya\ suggests the possible presence of a neutral gas cloud 
at a distance of 2.1 kpc from ID14, 
reached by the ionizing photons escaped from the system, where 
they can produce \lya\ fluorescence \citep[][]{MD16,beh14}.
The emerging spectral line profile apparently symmetric and narrow ($<40$\kms)  would also support the
fluorescence scenario. It is worth noting that ID14 shows 
optical Oxygen line equivalent widths and 
ratios similar to the ones of the Lyman continuum emitter 
discovered at z=3.2, named {\it Ion2}  (with confirmed $>50$\% escaping ionizing 
radiation, see \citealt{debarros16, vanz16b})\footnote{Though more massive ($<10^{9}$\msun) and 
larger ($\simeq 300$pc effective radius), {\it Ion2} shows an equivalent width
of the \oiiiv\ line $>1000$\AA\ rest-frame and a large O32 index ($>10$); where the O32 index is  (\oiiidoub) / (\oiidoub).}
and  some local extreme Green Peas sources (also Lyman continuum leakers, \citealt{izotov16,schaerer16}). 
A possible link between the large O32 index and the optically thin interstellar medium to ionizing radiation was
explored with photo-ionization models by  \citet[][]{jaskot13}.
The large value observed here (O32 $>10$) may indicate a density-bounded condition
(due to a deficient \oiidoub, \citealt{nakajima14}). This could be the case along some directions
(not necessarily along the line of sight), in which the opacity to ionizing photons is low. 
Interestingly, similar transverse cavities possibly optically thin in the Lyman continuum have been
observed in the local super star clusters, where expanding shells of gas 
(superbubbles) and blow-out regions can be produced by substantial mechanical feedback generated by 
massive super star clusters \citep[e.g., Mrk71,][]{james16,micheva17}. 

 In this scenario, the spatially offset fluorescent \lya\ nebula might reside into an elongated
 cavity directly connecting it with ID14.
 We would also expect a large escape fraction of  \lya\ photons along the same transverse direction \citep[][]{beh14}.

Alternatively, such \lya\ emission might be produced by additional faint/undetected sources. We identified
a faint object behind the diffuse \lya\ emission
(see the orange-dashed circle in Figure~\ref{muse}, panel C), detected in the HST bands 
with an average observed magnitude of $\simeq 30$.  
The faintness of the object prevents us from deriving any 
reliable photometric redshift or physical properties. We only note from Figure~\ref{orange} 
an apparent drop in the F435W and F606W bands when compared to the other redder bands.
If this object (or another undetected source) is responsible for the entire \lya\ diffuse emission, 
then its de-lensed magnitude would be $\gtrsim 34$ ($\mu \gtrsim 50$) with an EW(\lya)$>$450\AA\ 
rest-frame (where the magnification is the median value calculated within $1'' \times 1''$ centered on the object,
orange circle in Figure~\ref{orange}). This would correspond to an absolute magnitude of $M_{UV}=-11$
and the very large \lya\ equivalent width would imply the presence of unusual stellar
populations \citep[e.g.,][]{schaerer03, dijkstra07}. 
Since the redshift of this object is unconstrained and/or we don't know if other 
undetected sources were present, more specific analysis will need deeper imaging with future
instrumentation such as  JWST or ELT.

\section{Conclusions}
We presented a strongly lensed pair of super star-clusters at $z=3.2222$ separated
by $\sim 300$pc, with magnitudes $m\simeq30.6$ and effective radius of 30pc each. 
Both show high-ionization  lines like \civ, \heii, \oiiiuv\ and \ciiidoub\ with S/N ratios
ranging between 2 and 20 and  \hb, \oiiidoub\ optical lines with S/N=$4-31$, all of them
showing narrow velocity dispersions, $\sigma_v<50$\kms\ (derived from MUSE) or $\sigma_v \lesssim 20$\kms\ 
(derived from X-Shooter). The presence of high-ionization lines (like \heii) and the physical
properties inferred from the observed colors (SED-fitting) suggest young stars are present 
($<10$ Myr), as well as a relatively evolved stellar component ($<100$ Myr). A moderate
dust attenuation (E(B-V) $\simeq 0.06$) is also consistent with the observed colors.
The inferred stellar masses are of a few $10^{6}$ \msun\ with a metallicity 1/10 solar.

The main results can be summarized as follow: 

\begin{enumerate}

\item{Despite the fact that the above properties are often observed in \lya\ emitting galaxies such as Green Peas,
 \citep[e.g.,][]{henry15}, ID14 does not show any \lya\ emission aligned with the detected stellar
continuum and ultraviolet metal lines, in all the observed multiple images ID14a,b,c. In particular
the ratio \civ\ / \lya\ $\gtrsim 15$ is uncommon, and it is the first case confirmed at high redshfit and 
very low-luminosity regimes as discussed here. A relatively low velocity outflow, the presence of a
screen of neutral gas and the presence of an (even moderate) amount of dust are all ingredients that
may explain the \lya\ attenuation.} 

\item{A spatially offset and strongly magnified  \lya\  emission is detected at $\sim 2$kpc transverse 
from  ID14. There are two possible explanations: (A)  an induced fluorescence
in a nearby neutral gas cloud through a local and transverse escape of ionizing radiation from ID14
(this would also imply a transverse escape of \lya\ photons along the same route, see, e.g., \citealt{beh14}) 
and/or (B) an in-situ star formation activity of an object barely (or not) detected in the deep HST images,
with de-lensed $M_{UV} \simeq -11$ ($m>34$) and EW(\lya)$ > 450$\AA\ rest-frame, the latter, however,
would imply unusual stellar populations \citep[e.g.,][]{can12, dijkstra07}.} 

\end{enumerate}

The double lens effect (see Sect.~2),
in which the observed flux ratios among the multiple images allow us to derive a relatively small error
in a high magnification regime, $\mu=40 \pm 7$, offers the opportunity to anticipate the expected 
future ELT capabilities in terms of the spatial resolution  and magnitudes in the not lensed 
fields \citep[see also,][]{christ12, vanz16a}.
In particular, besides the detected pair of super star clusters (components ``1,2''), additional sub-structures
have been identified possibly representing single \hii\ regions and/or candidate proto-globular
clusters \citep[][]{vanz17b}. More investigation is needed to fully characterize these systems.
The spatial resolution (3-7 mas) achievable with the ELT instrumentation in conjunction with 
strong lensing (as in this case) will offer the opportunity to spatially resolve details of 1-2 pc at $z>3$. 
It is also worth reporting what integration time would be needed with MUSE to achieve the same 
S/N ratios of UV metal lines for a ID14-like not lensed object. The most prominent among
them is the \civblue\ with de-lensed flux of $2.2\times 10^{-19}$ \ergscm\ that would require
$\simeq 7200$ hours integration time to reproduce the same S/N=17 reported here 
(or 240 hours for a S/N=3, with typical observing conditions). 
Again, the optical spectrum reported in this work offer a first glance on the future ELT-like spectra
of similar objects in the non-lensed fields, attainable with several hours integration time.
 
Finally, we demonstrated (at least for this faint system) that the \lya\ / \civalone\ ratio
often reported in literature \citep[e.g.,][]{stark14,stark15,mainali17} 
 or any line ratio that includes the \lya\ flux (e.g., Figure 7 in \citealt{shibuya17}) can be
affected by large line-of-sight variation of the \lya\ visibility also in low-luminosity regimes, 
therefore weakening its use as a diagnostic feature \citep[e.g.,][]{smit17}.

\acknowledgments
We thank the referee for useful comments that improved the manuscript.
We thank K. Schmidt, L. Mas-Ribas and R. Amorin for useful discussion.
C.G. acknowledges support by VILLUM FONDEN Young
Investigator Programme through grant no. 10123.  K.C. acknowledges
funding from the European Research Council through the award of the
Consolidator Grant ID 681627-BUILDUP.  LC is supported by Grant DFF 4090-00079.
Based on observations collected
at the European Southern Observatory for Astronomical research in the
Southern Hemisphere under ESO programmes P095.A-0653, P094.A-0115 (B)
and ID 094.A-0525(A). MM, AM and PR acknowledge the financial support
from PRIN-INAF 2014 1.05.01.94.02.


\begin{thebibliography}{}
\bibitem[Alavi et al.(2016)]{alavi16} Alavi, A., Siana, B., Richard, J., et al.\ 2016, arXiv:1606.00469
\bibitem[Amor{\'{\i}}n et al.(2015)]{amorin15} Amor{\'{\i}}n, R., P{\'e}rez-Montero, E., Contini, T., et al.\ 2015, \aap, 578, A105
\bibitem[Amor{\'{\i}}n et al.(2017)]{amorin17} Amor{\'{\i}}n, R., Fontana, A., P{\'e}rez-Montero, E., et al.\ 2017, Nature Astronomy, 1, 0052
\bibitem[Annibali et al.(2015)]{annibali15} Annibali, F., Tosi, M., Pasquali, A., et al.\ 2015, \aj, 150, 143
\bibitem[Atek et al.(2009)]{atek09} Atek, H., Schaerer, D., \& Kunth, D.\ 2009, \aap, 502, 791 
\bibitem[Bacon et al.(2010)]{bacon10} Bacon, R., Accardo, M., Adjali, L., et al.\ 2010, \procspie, 7735, 773508
\bibitem[Behrens et al.(2014)]{beh14} Behrens, C., Dijkstra, M., \& Niemeyer, J.~C.\ 2014, \aap, 563, A77 
\bibitem[\protect\citeauthoryear{Brocklehurst}{1971}]{brocklehurst71} Brocklehurst M., 1971, MNRAS, 153, 471 
\bibitem[Bouwens et al.(2017)]{bouwens17} Bouwens, R.~J., Illingworth, G.~D., Oesch, P.~A., et al.\ 2016, arXiv:1608.00966  
\bibitem[Bruzual \& Charlot(2003)]{BC03} Bruzual, G., \& Charlot, S.\ 2003, \mnras, 344, 1000 
\bibitem[Caminha et al.(2016a)]{cam16a} Caminha, G.~B., Grillo, C., Rosati, P., et al.\ 2016, \aap, 587, A80             
\bibitem[Caminha et al.(2016b)]{cam16b} Caminha, G.~B., Karman, W., Rosati, P., et al.\ 2016, \aap, 595, A100   
\bibitem[Caminha et al.(2016c)]{cam16c} Caminha, G.~B., Grillo, C., Rosati, P., et al.\ 2016, arXiv:1607.03462    
\bibitem[Cantalupo et al.(2012)]{can12} Cantalupo, S., Lilly, S.~J., \& Haehnelt, M.~G.\ 2012, \mnras, 425, 1992
\bibitem[Cardamone et al.(2009)]{cardamone09} Cardamone, C., Schawinski, K., Sarzi, M., et al.\ 2009, \mnras, 399, 1191 
\bibitem[Castellano et al.(2016b)]{castellano16b} Castellano, M., Amor{\'{\i}}n, R., Merlin, E., et al.\ 2016, \aap, 590, A31          
\bibitem[Castellano et al.(2016a)]{castellano16a} Castellano, M., Dayal, P., Pentericci, L., et al.\ 2016, \apjl, 818, L3    
\bibitem[Christensen et al.(2012)]{christ12} Christensen, L., Laursen, P., Richard, J., et al.\ 2012, \mnras, 427, 1973
\bibitem[Dijkstra \& Wyithe(2007)]{dijkstra07} Dijkstra, M., \& Wyithe, J.~S.~B.\ 2007, \mnras, 379, 1589
\bibitem[Duval et al.(2016)]{duval16} Duval, F., {\"O}stlin, G., Hayes, M., et al.\ 2016, \aap, 587, A77
\bibitem[de Barros et al.(2016)]{debarros16} de Barros, S., Vanzella, E., Amor{\'{\i}}n, R., et al.\ 2016, \aap, 585, A51
\bibitem[Erb et al.(2014)]{erb14} Erb, D.~K., Steidel, C.~C., Trainor, R.~F., et al.\ 2014, \apj, 795, 33
\bibitem[Erb et al.(2016)]{erb16} Erb, D.~K., Pettini, M., Steidel, C.~C., et al.\ 2016, \apj, 830, 52 
\bibitem[Feltre et al.(2016)]{feltre16} Feltre, A., Charlot, S., \& Gutkin, J.\ 2016, \mnras, 456, 3354 
\bibitem[Fosbury et al.(2003)]{fosbury03} Fosbury, R.~A.~E., Villar-Mart{\'{\i}}n, M., Humphrey, A., et al.\ 2003, \apj, 596, 797 
\bibitem[Furlanetto et al.(2005)]{furlanetto05} Furlanetto, S.~R., Schaye, J., Springel, V., \& Hernquist, L.\ 2005, \apj, 622, 7 
\bibitem[Grillo et al.(2016)]{grillo16} Grillo, C., Karman, W., Suyu, S.~H., et al.\ 2016, \apj, 822, 78
\bibitem[Gronke \& Dijkstra(2014)]{gronke14} Gronke, M., \& Dijkstra, M.\ 2014, \mnras, 444, 1095 
\bibitem[Gutkin et al.(2016)]{gutkin16} Gutkin, J., Charlot, S., \& Bruzual, G.\ 2016, \mnras, 462, 1757
\bibitem[Henry et al.(2015)]{henry15} Henry, A., Scarlata, C., Martin, C.~L., \& Erb, D.\ 2015, \apj, 809, 19 
\bibitem[Herenz et al.(2016)]{herenz16} Herenz, E.~C., Gruyters, P., Orlitova, I., et al.\ 2016, \aap, 587, A78 
\bibitem[Holden et al.(2016)]{holden16} Holden, B.~P., Oesch, P.~A., Gonz{\'a}lez, V.~G., et al.\ 2016, \apj, 820, 73 
\bibitem[Izotov et al.(2016)]{izotov16} Izotov, Y.~I., Schaerer, D., Thuan, T.~X., et al.\ 2016, \mnras, 461, 3683 
\bibitem[James et al.(2016)]{james16} James, B.~L., Auger, M., Aloisi, A., Calzetti, D., \& Kewley, L.\ 2016, \apj, 816, 40
\bibitem[Jaskot \& Oey(2013)]{jaskot13} Jaskot, A.~E., \& Oey, M.~S.\ 2013, \apj, 766, 91 
\bibitem[Karman et al.(2017)]{karman17} Karman, W., Caputi, K.~I., Caminha, G.~B., et al.\ 2017, A\&A, 599, 28
\bibitem[Kehrig et al.(2015)]{kehrig15} Kehrig, C., V{\'{\i}}lchez, J.~M., P{\'e}rez-Montero, E., et al.\ 2015, \apjl, 801, L28
\bibitem[Koekemoer et al.(2014)]{koekemoer14} Koekemoer, A.~M., Avila, R.~J., Hammer, D., et al.\ 2014, American Astronomical Society Meeting Abstracts \#223, 223, 254.02 
\bibitem[Lotz et al.(2014)]{lotz14} Lotz, J., Mountain, M., Grogin, N.~A., et al.\ 2014, American Astronomical Society Meeting Abstracts \#223, 223, 254.01 
\bibitem[Lotz et al.(2016)]{lotz16} Lotz, J.~M., Koekemoer, A., Coe, D., et al.\ 2016, arXiv:1605.06567
\bibitem[Mainali et al.(2017)]{mainali17} Mainali, R., Kollmeier, J.~A., Stark, D.~P., et al.\ 2016, arXiv:1611.07125
\bibitem[Mas-Ribas \& Dijkstra(2016)]{MD16} Mas-Ribas, L., \& Dijkstra, M.\ 2016, \apj, 822, 84
\bibitem[Meneghetti et al.(2017)]{meneghetti17} Meneghetti, M., Natarajan, P., Coe, D., et al.\ 2016, arXiv:1606.04548 
\bibitem[Merlin et al.(2016)]{merlin16} Merlin, E., Amor{\'{\i}}n, R., Castellano, M., et al.\ 2016, \aap, 590, A30 
\bibitem[Micheva et al.(2017)]{micheva17} Micheva, G., Oey, M.~S., Jaskot, A.~E., \& James, B.~L.\ 2017, arXiv:1704.01678 
\bibitem[Nakajima \& Ouchi(2014)]{nakajima14} Nakajima, K., \& Ouchi, M.\ 2014, \mnras, 442, 900 
\bibitem[Ogrean et al.(2015)]{ogrean15} Ogrean, G.~A., van Weeren, R.~J., Jones, C., et al.\ 2015, \apj, 812, 153 
\bibitem[Peng et al.(2010)]{peng10} Peng, C.~Y., Ho, L.~C., Impey, C.~D., \& Rix, H.-W.\ 2010, \aj, 139, 2097 
\bibitem[Pilyugin et al.(2006)]{pilyugin06} Pilyugin, L.~S., V{\'{\i}}lchez, J.~M., \& Thuan, T.~X.\ 2006, \mnras, 370, 1928 
\bibitem[P{\'e}rez-Montero \& Amor{\'{\i}}n(2017)]{perez17} P{\'e}rez-Montero, E., \& Amor{\'{\i}}n, R.\ 2017, \mnras,
\bibitem[Rauch et al.(2016)]{rauch16} Rauch, M., Becker, G.~D., \& Haehnelt, M.~G.\ 2016, \mnras, 455, 3991 
\bibitem[Rhoads et al.(2014)]{rhoads14} Rhoads, J.~E., Malhotra, S., Richardson, M.~L.~A., et al.\ 2014, \apj, 780, 20
\bibitem[Schaerer(2003)]{schaerer03} Schaerer, D.\ 2003, \aap, 397, 527 
\bibitem[Schaerer et al.(2016)]{schaerer16} Schaerer, D., Izotov, Y.~I., Verhamme, A., et al.\ 2016, \aap, 591, L8
\bibitem[Schmidt et al.(2017)]{schmidt17} Schmidt, K.~B., Huang, K.-H., Treu, T., et al.\ 2017, arXiv:1702.04731
\bibitem[Shibuya et al.(2017)]{shibuya17} Shibuya, T., Ouchi, M., Harikane, Y., et al.\ 2017, arXiv:1705.00733 
\bibitem[Smit et al.(2017)]{smit17} Smit, R., Swinbank, A.~M., Massey, R., et al.\ 2017, \mnras,
\bibitem[Smith et al.(2016)]{smith16} Smith, L.~J., Crowther, P.~A., Calzetti, D., \& Sidoli, F.\ 2016, \apj, 823, 38
\bibitem[Stark et al.(2014)]{stark14} Stark, D.~P., Richard, J., Siana, B., et al.\ 2014, \mnras, 445, 3200
\bibitem[Stark et al.(2015)]{stark15} Stark, D.~P., Walth, G., Charlot, S., et al.\ 2015, \mnras, 454, 1393
\bibitem[Stark et al.(2017)]{stark17} Stark, D.~P., Ellis, R.~S., Charlot, S., et al.\ 2017, \mnras, 464, 469
\bibitem[Steidel et al.(2014)]{steidel14} Steidel, C.~C., Rudie, G.~C., Strom, A.~L., et al.\ 2014, \apj, 795, 165 
\bibitem[Vanzella et al.(2014)]{vanz14} Vanzella, E., Fontana, A., Zitrin, A., et al.\ 2014, \apjl, 783, L12   
\bibitem[Vanzella et al.(2017a)]{vanz17a} Vanzella, E., Balestra, I., Gronke, M., et al.\ 2017, \mnras, 465, 3803 
\bibitem[Vanzella et al.(2016b)]{vanz16b} Vanzella, E., de Barros, S., Vasei, K., et al.\ 2016, \apj, 825, 41            
\bibitem[Vanzella et al.(2016a)]{vanz16a} Vanzella, E., De Barros, S., Cupani, G., et al.\ 2016, \apjl, 821, L27     
\bibitem[Vanzella et al.(2017b)]{vanz17b} Vanzella, E., Calura, F., Meneghetti, M., et al.\ 2017, \mnras, 467, 4304 
\bibitem[Verhamme et al.(2012)]{verhamme12} Verhamme, A., Dubois, Y., Blaizot, J., et al.\ 2012, \aap, 546, A111 
\bibitem[Weidinger et al.(2005)]{weidinger05} Weidinger, M., M{\o}ller, P., Fynbo, J.~P.~U., \& Thomsen, B.\ 2005, \aap, 436, 825 
\bibitem[Zheng \& Wallace(2014)]{zheng14} Zheng, Z., \& Wallace, J.\ 2014, \apj, 794, 116 
\bibitem[Zitrin et al.(2013)]{zitrin13} Zitrin, A., Meneghetti, M., Umetsu, K., et al.\ 2013, \apjl, 762, L30


\end{thebibliography}
\end{document}